\documentclass[a4paper,11pt]{article}
\pdfoutput=1
\usepackage{jinstpub} 
\usepackage{lineno}
\title{\boldmath Progress report on the online processing upgrade at the NA62 experiment}

\author[b]{R.~Ammendola}
\author[a]{A.~Biagioni}
\author[a]{A.~Ciardiello}
\author[a]{P.~Cretaro}
\author[a]{O.~Frezza}
\author[e,1]{G.~Lamanna\note{also at INFN, Sezione di Pisa, Italy}}
\author[a]{F.~Lo~Cicero}
\author[a]{A.~Lonardo}
\author[a]{M.~Martinelli}
\author[f]{R.~Piandani}
\author[a]{L.~Pontisso}
\author[d,2]{M.~Raggi\note{also at INFN, Sezione di Roma, Italy}}
\author[a]{F.~Simula}
\author[c]{D.~Soldi}
\author[a,3]{M.~Turisini\note{Corresponding author.}}
\author[a]{P.~Vicini}

\affiliation[a]{INFN, Sezione di Roma, Italy}
\affiliation[b]{INFN, Sezione di Roma Tor Vergata, Italy}
\affiliation[c]{INFN, Sezione di Torino, Italy}
\affiliation[d]{Università Sapienza di Roma, Italy}
\affiliation[e]{Università di Pisa, Italy}
\affiliation[f]{University of Chinese Academy of Science, Beijing, China}

\emailAdd{matteo.turisini@roma1.infn.it}

\abstract{
A new FPGA-based low-level trigger processor has been installed at \rev{the} NA62 experiment.
It is intended to extend the features of its predecessor \rev{due} to a faster interconnection technology and \rev{additional} logic resources available on the \rev{new} platform.
With the aim of improving trigger selectivity and exploring new architectures for complex trigger computation, a GPU system has been developed and a neural network on FPGA is in progress.
They both process data streams from the Ring Imaging Cherenkov detector of the experiment to extract in real time high level features for the trigger logic.
Description of the systems, latest developments and design flows are reported in this paper.
}


\proceeding{Topical Workshop on Electronics for Particle Physics -TWEPP2021 \\ 20 - 24 September, 2021 \\ online}

\usepackage{xspace}
\newcommand{\lzerotp}{L0TP\xspace}
\newcommand{\lzerotpp}{L0TP+\xspace}

\newcommand{\rev}[1]{\textcolor{black}{#1}}

\begin{document}
\maketitle
\flushbottom
\section{Introduction}
NA62 is a Kaon physics experiment located in the north area of the SPS complex at CERN.
It adopts a novel decay-in-flight technique to record ultra rare decays of K mesons from a 75 GeV/c positively charged hadron beam.  
The apparatus \rev{can detect} kaons and related decay products \rev{using} a wide range of detectors distributed along a 270 meter\rev{s} long beamline.
A proximity focusing Ring Imaging Cherenkov (RICH) is used to increase the muon/pion separation in the final state \rev{in addition to} calorimetry and kinematic detectors.
The signals of interest can \rev{occur with a frequency} 10 orders of magnitude less than the background, \rev{requiring} a beam of high intensity \rev{coupled} with good trigger selectivity.
The nominal intensity is 750~MHz for the hadron beam with a 10 MHz particles rate on the RICH.

The online data selection is operated by a fully digital multilevel trigger together with a network-based data acquisition system (TDAQ). 
A common infrastructure distributes clock, backpressure and reference signals to all detectors.
The lowest level trigger is called  Level Zero (L0) and it is implemented on an Altera Stratix IV FPGA.
The task of L0 is to produce trigger decisions based on minimal information (called trigger \emph{primitives}) received from a subgroup of fast detectors with a latency less than one millisecond.

Our work has focused on the hardware upgrade of the L0 trigger processor (\lzerotpp) and on the development of two innovative modules \rev{that improve the information encoded in the RICH primitives}.
At present the RICH primitives encode only hit multiplicity, while the proposed systems can calculate ring related quantities relying on high throughput trackless seeded algorithms.
The paper is organized as follows:
section~\ref{sec:gpurich} describes a GPU-based ring reconstruction system currently under finalization at the experiment (GPURICH); 
section~\ref{sec:l0tpp} presents \lzerotpp hardware \rev{which} has been recently installed at the experiment and is currently under comparative validation test;
section~\ref{sec:rinngs} introduces a preliminary neural network classifier on FPGA for real time inference over \rev{the} RICH data stream (RiNNgs).

\section{GPURICH}\label{sec:gpurich}
\begin{figure}[b]
\centering 
\includegraphics[width=.36\textwidth]{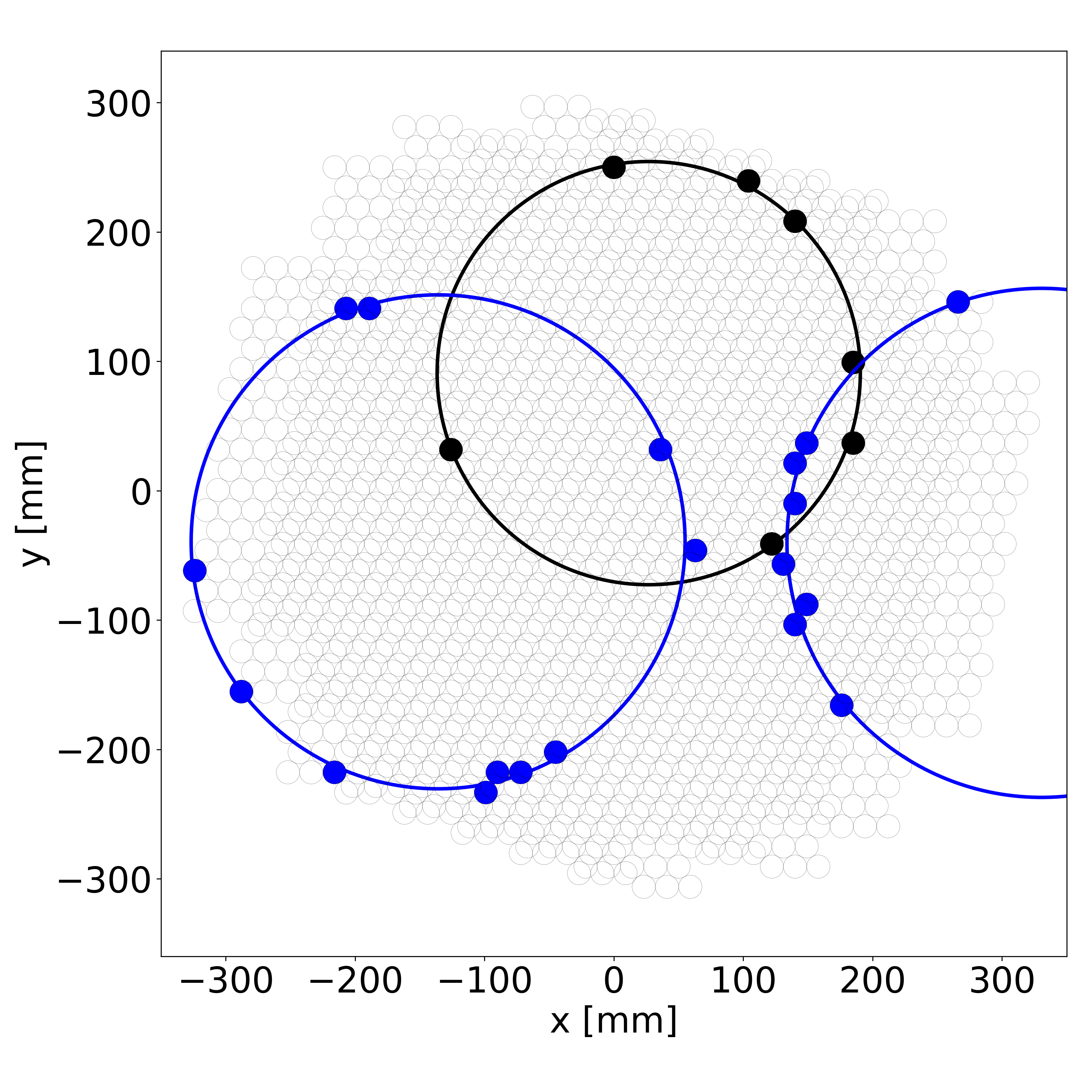}
\qquad
\includegraphics[width=.55\textwidth]{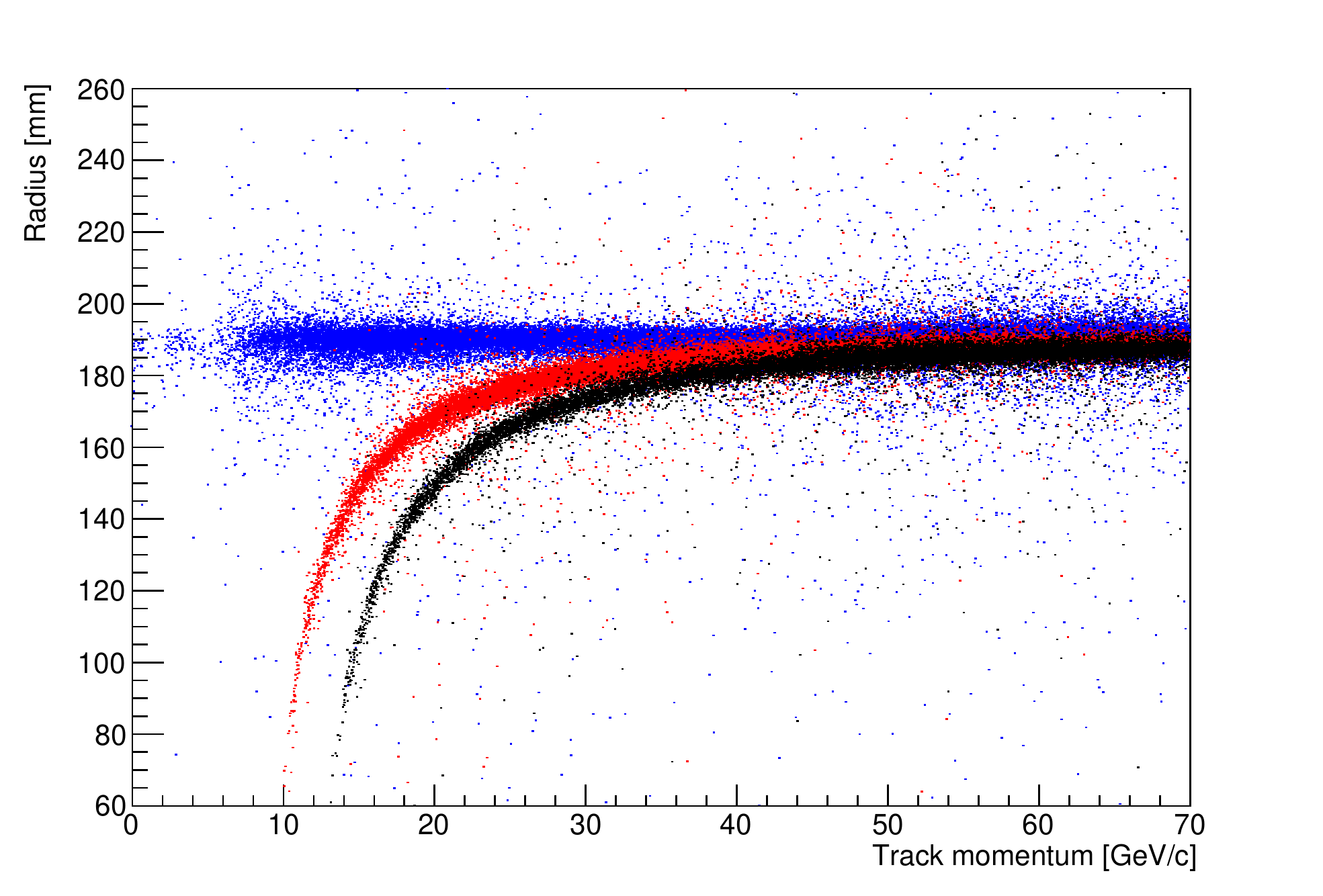}
\caption{\label{fig:eventdisplay}
Left: A hit pattern on the RICH array with reconstructed rings geometry superimposed. 
Right: Ring radius versus particle momentum for 130k events.
Electrons in blue,  muons in red and pions in black. 
Particles are identified offline combining RICH, tracking and calorimetry data.}
\end{figure}


The RICH detector of NA62 was designed for pion/muon separation in the range 15 to 35 GeV/c.
A typical performance plot and an event with three charged particles are shown in figure~\ref{fig:eventdisplay} as examples of its particle identification (PID) capability.
The RICH  electronics consist\rev{s} \rev{of} an array of 1952 photomultiplier tubes (PMT), 64 frontend cards and 4 readout boards in charge of data digitization, timestamping, buffering and networking.
In the baseline TDAQ configuration the RICH primitives are generated by means of a fifth readout board\rev{.}
\rev{The signal sent to L0 is derived from a simple OR of} 8 adjacent PMTs, \rev{resulting in a trigger logic that is} based on hit multiplicity at low granularity and no ring information.
This solution is motivated by an intrinsically high time precision at the level of 70 picoseconds \cite{Anzivino:RICH2020}, \rev{and }practical absence of random coincidences.
\rev{However, it does not exploit any PID capability of the detector, which is fully used during offline analysis instead.}

\rev{A new } GPU-based system for online ring fitting and physics-related primitive generation, called GPURICH, has \rev{therefore} been proposed and installed.
It operates on the full detector granularity and leverages the FPGA-based custom network interface called NaNet~\cite{Ammendola:twepp2014} to transfer data with low and controlled latency from the detector to the GPU memory.
Primitives are then computed using a histogram based ring detection and a standard ring fitting algorithm.
The adopted boards are a Terasic DE5 \rev{with} an Altera Stratix-V FPGA for NaNet, and a NVIDIA Pascal P100 GPU for the ring geometry reconstruction.
The two boards communicate directly through GPUdirect RDMA protocol on PCIe bus.
Data from \rev{the} detector to the trigger processor are sent using eight 1GbE links aggregated through a switch to a 10GbE \rev{link}. 
Since GPURICH provides physics related quantities like the number of particles and the ring radius information already at L0, it can participates in the trigger decision as a smart additional detector.

During NA62 Run1 (in 2018) GPURICH was installed in the experimental hall and the synchronization with other detectors generating primitives was validated. 
A total latency well below one millisecond has been \rev{measured}, with an average processing time of 130 ns per event~\cite{Cretaro:twepp2018}.
\rev{For} Run 2 (2021-2022) the efforts are devoted to specialize the algorithms to electron identification.

\section{\lzerotpp} \label{sec:l0tpp}
\begin{figure}[b]
\centering 
\includegraphics[trim=4cm 4cm 8cm 4cm,clip,width=.7\textwidth]{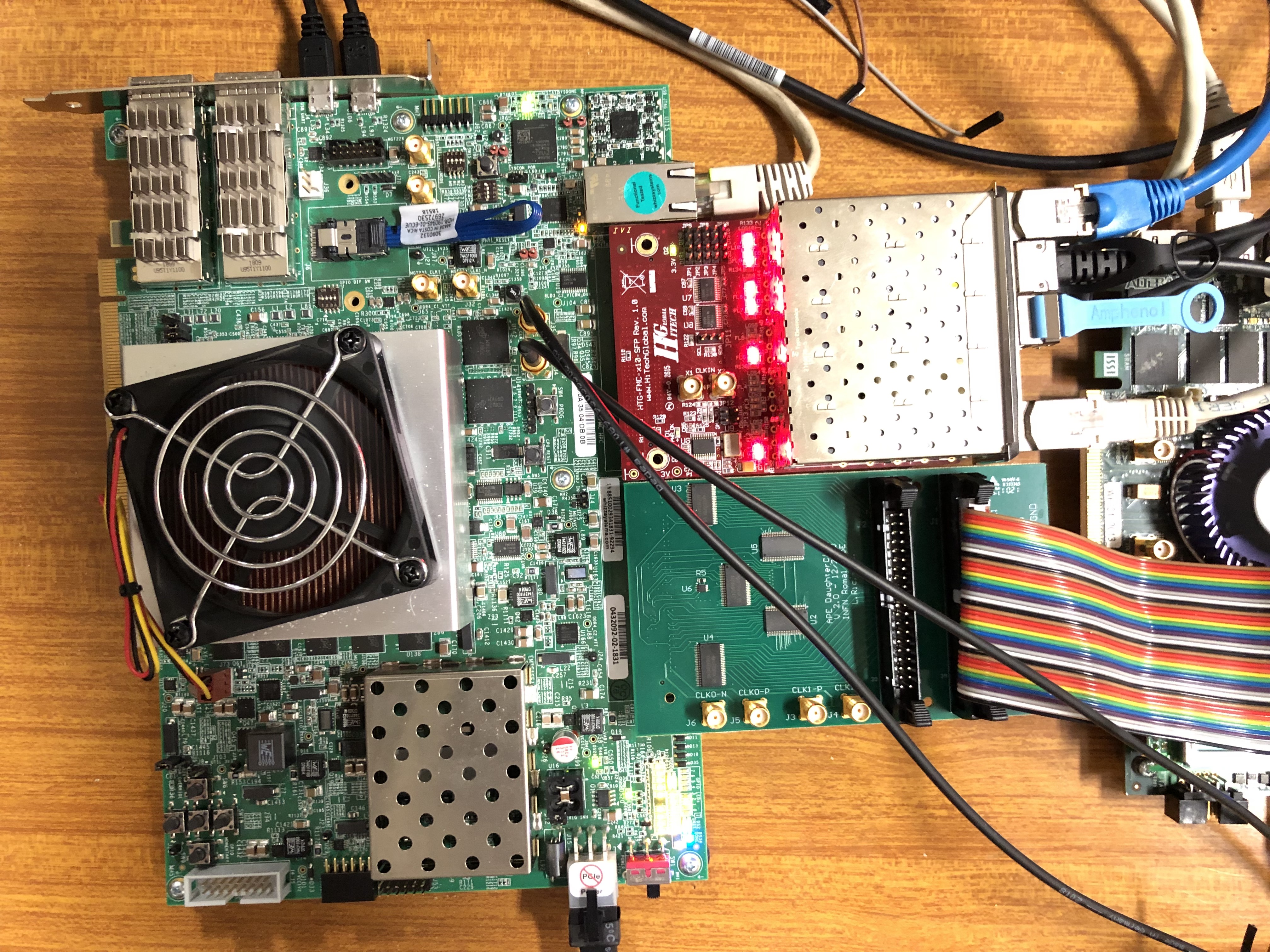}
\caption{\label{fig:testbed} \lzerotpp assembly: Xilinx Virtex UltraScale+ FPGA VCU118 board with 2 mezzanine cards to interface with detectors, PC farm and control signals from NA62 infrastructure.
}
\end{figure}

The Level Zero Trigger Processor (\lzerotp) of NA62 is entirely hardware based and implemented on \rev{an} FPGA \cite{Ammendola:nima2019}.
It receives trigger \emph{primitives} from a subset of detectors, operates logical coincidences between them \rev{according} to user defined masks and sends \rev{the} trigger decision to the NA62 common infrastructure depending on a programmable downscaling factor for each mask.
A proximity PC is used to interface with the Run Control of the experiment, debug and \rev{collect data} for offline trigger analysis.
Current implementation is a on a Terasic DE4 board housing a Stratix-IV FPGA device and eight 1GbE links. 

The  hardware and logic resources available on \rev{the current} \lzerotp cannot sustain a beam intensity increase nor additional computations to improve selectivity.
To overcome \rev{these} limitations a more recent platform \rev{has been} adopted and the new system is called \lzerotpp \cite{Ammendola:chep2019}. 
An illustration of the assembly based on \rev{a} Xilinx Virtex UltraScale+ FPGA VCU118 board is \rev{shown} in figure \ref{fig:testbed}.
In parallel with a test bench for characterization and \rev{debugging} at \rev{the} INFN laboratories, \rev{the} \lzerotpp has been installed in the experimental hall and validation is ongoing to replicate the performance 
obtained so far 
\rev{of the previous system}
before adding new features.

A 10GbE channel has been already added to support \rev{an output rate of} up to 10MHz and a microcontroller soft IP has been embedded in the new FPGA device for a smoother integration with the Run Control.
Currently occupied resources are $23\%$ of IO pins, $19\%$ Gigabit transceivers, $19\%$ BRAM, $8\%$ of LUT, $6\%$ of registers and less than $1\%$ of DSP blocks.
The \rev{additional}  resources allows also for new trigger algorithms to be implemented as described in the next section. 

\section{RiNNgs} \label{sec:rinngs}

\begin{figure}[b]
\centering
\includegraphics[width=.42\textwidth]{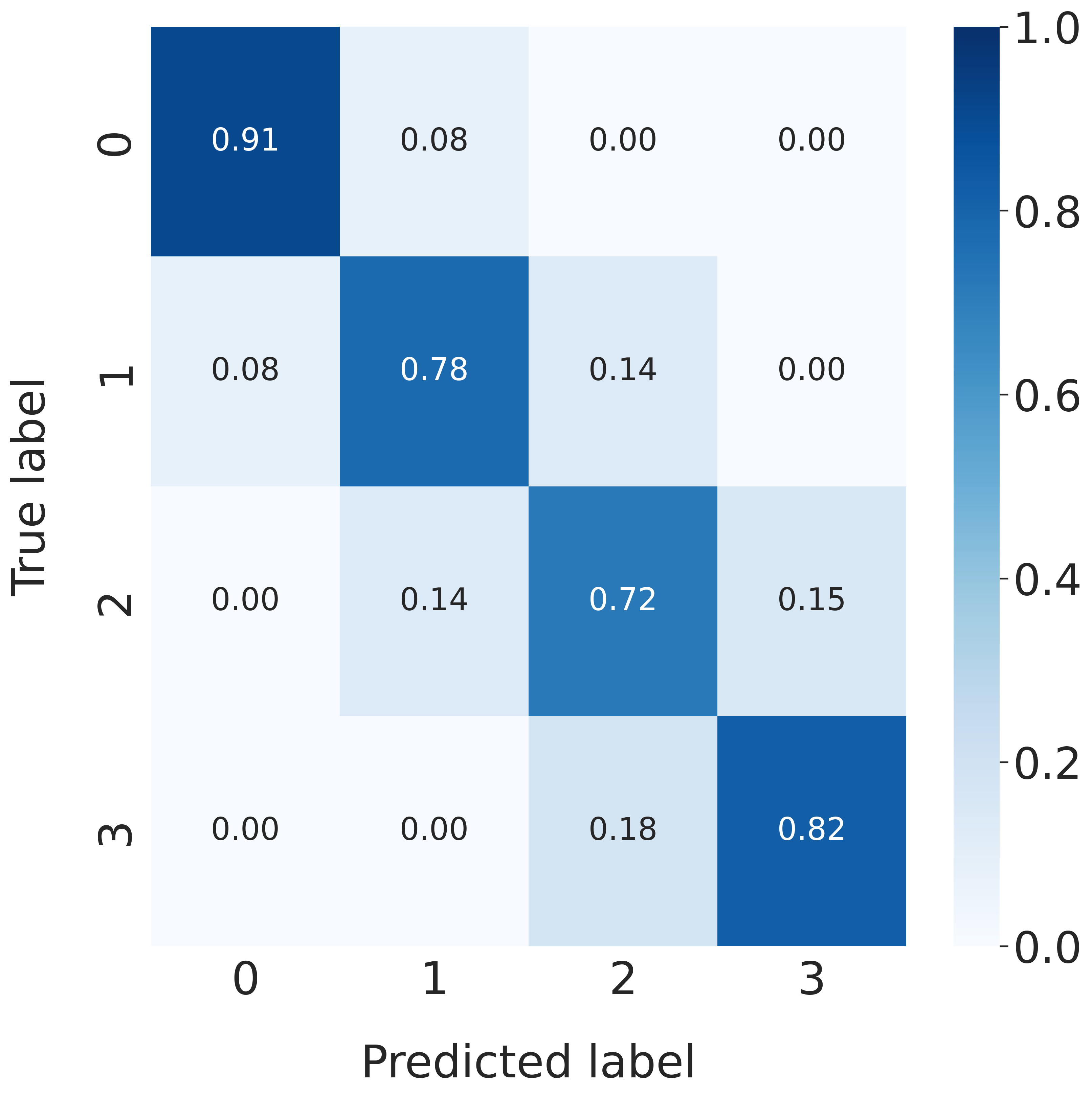}
\qquad
\includegraphics[width=.45\textwidth]{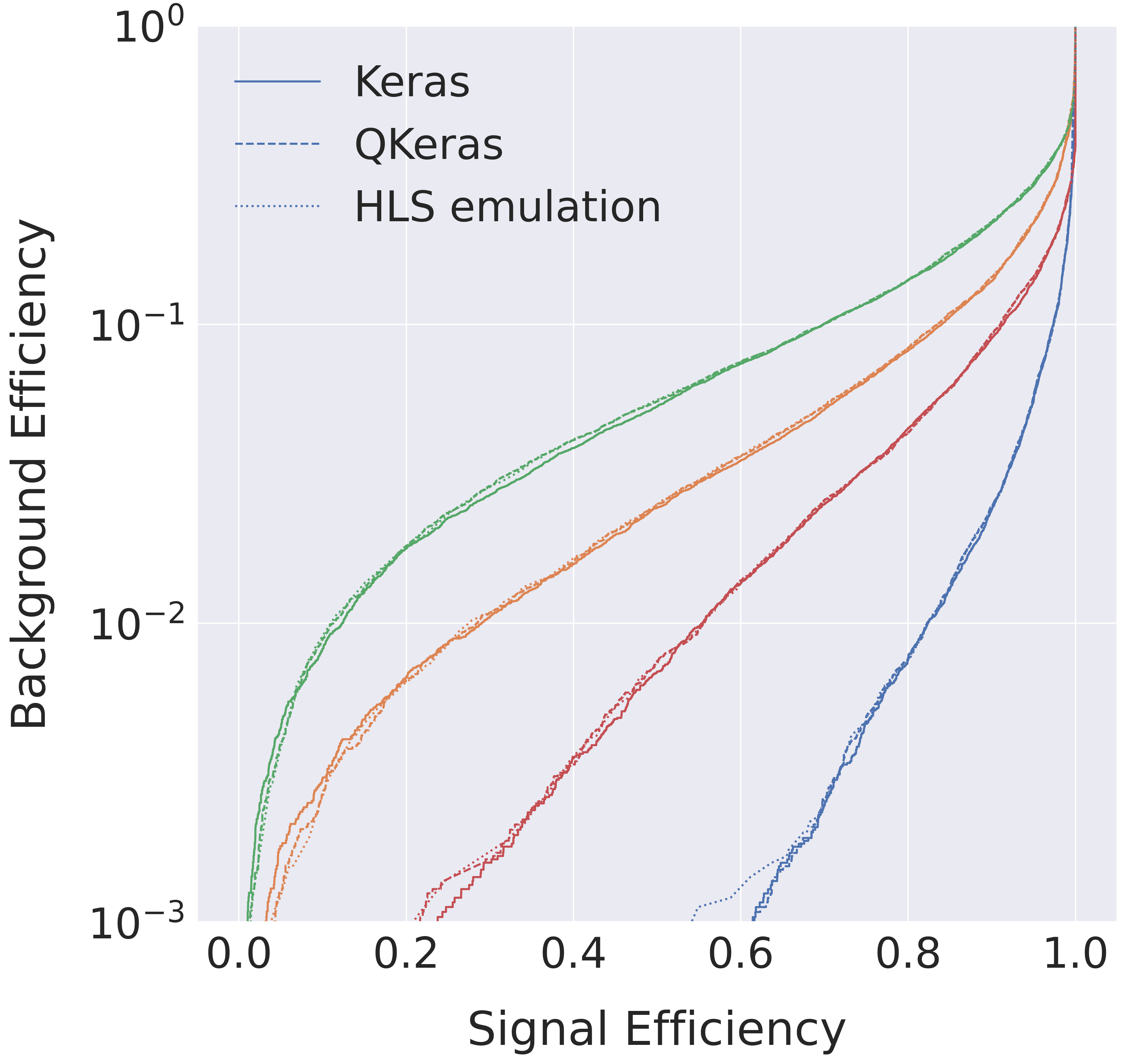}
\caption{\label{fig:rinngsPerformance} 
\rev{RiNNgs particle counting performance, the model is trained to predict the number of charged particles (label) in each event.
Left: Confusion matrix result for validation dataset (8k events). 
Right: Simulation results for the 4 output labels i.e. rings: 0 (blue), 1 (orange), 2 (green), 3 or more (red).
Reducing the numerical precision (QKeras) does not degrade the floating point (Keras) performance.
This result is confirmed using an emulated version of the FPGA design (HLS emulation).}
}
\end{figure}
An event classifier for the RICH detector compatible with L0 timing requirements has been developed using machine learning techniques and high level synthesis tools for FPGA design.
This is motivated by a potential benefits for the trigger selectivity given by the full exploitation of the Cherenkov signals at the online level and is an example of how offline processing can be turned online 
\rev{due to the latest developments in hardware and software}.
As a first case study, the task of counting the number of rings was considered
and even if ring features extraction can be a trivial task for modern deep neural networks, the 10 MHz particle flux on the RICH still represents a challenge.

\rev{The classifier is called Rings Neural Network (RiNNgs)} and is composed \rev{of} 3 fully connected layers with respectively 64, 16 and 4 output neurons activated by ReLu functions.
It takes the event hit list as input (normalized by the number of front end channels) and produces the probability of having 0,1,2,3 or more rings in output (labels).
The dataset consists of 80k events per label and was obtained within \rev{the} NA62 \rev{a}nalysis framework \rev{using} experimental data and track seeded ring reconstruction.

Figure \ref{fig:rinngsPerformance} presents the inference performance after training in the standard TensorFlow framework.
\rev{For } the confusion matrix \rev{(fig. \ref{fig:rinngsPerformance}, left)} 
\rev {the} average accuracy performance is about $80\%$, satisfactory for \rev{the} L0 \rev{trigger}.
Moreover\rev{,} some trigger conditions could tolerate an overestimation of the number of particles\rev{,} rising the efficiency above $80\%$ for all the cases. 
On the right, the receiver operating characteristics \rev{are} shown for the four labels.
The curve can be used to \rev{optimize} the trigger for high efficiency or high purity.

The corresponding FPGA design is obtained using the HLS4ML Python package \cite{Duarte:2018ite} in combination with the Xilinx Vivado High Level Synthesis (HLS) tool.
The target board is the VCU118, since the \lzerotpp will be the system were RiNNgs is to be deployed.  
As shown in table~\ref{tab:hls} the design occupies a modest fraction of the available resources, \rev{allowing the neural network and the trigger processor to be placed on the same FPGA.} 
Table~\ref{tab:hls} \rev{also shows the HLS estimate for the} timing \rev{characteristics} which is compatible with a 10 MHz throughput.
After an initial 18 bits fixed point data representation (\emph{baseline}, 8 bits of integer part) a reduced precision (\emph{quantized}) attempt was made\rev{,} obtaining \rev{the} same performance with less DSP resources and better timing (see table~\ref{tab:hls} for details).
A necessary step to achieve this goal was to perform part of the training phase using a Python library called QKeras \cite{Coelho:2020zfu} that supports arbitrary quantization datatypes.

The design flow proved to be very effective for the fast implementation of \rev{a} neural network on FPGA.
However the following two \rev{limitations prevent} at the moment a better implementation: a) a limit of 4096 cycles on the unrolling factor for the pipelined design in Vivado HLS, and b) unsupported binary representation of the variables in QKeras (only weights and biases can be quantized).
Attempts are ongoing to overcome these limitations, both by editing manually the C++ code generated by the HLS4ML library and by migrating to the more recent Vitis HLS platform.

\begin{table}[]
\centering
\caption{\label{tab:hls} HLS report summary using different arithmetic representations.
\rev{The Instantiation Interval (II) is the minimum time distance between consecutive events to be processed. With a conservative estimation of 10 nanoseconds clock period the II results in 100 nanoseconds, thus compatible with the 10 MHz throughput requirement.}
The quantization is determined from the numerical range of the weights and biases after an initial training phase. The results showed are obtained with 7 and 9 bits fixed point representation with 1 bit dedicated to the integer part.}
\smallskip
\begin{tabular}{|c|cccc|ccc|}
\hline
          &BRAM  & DSP  &FF    & LUT  &  II            &LATENCY       & FMAX\\
          &[\%]  & [\%] &[\%]  & [\%] &[clock ticks] & [clock ticks]& [MHz]\\
\hline
baseline  & 1    & 10   & $<1$    & 3    & 10            & 40           & 100\\
quantized & 0    &  6   & $<1$    & 15   & 10            & 20           & 150\\
\hline
\end{tabular}
\end{table}
\section{Conclusions}\label{sec:conclusions}


The new low level trigger processor of NA62 is currently under test.
With 10 GbE support and \rev{significantly more logic,} it will support \rev{the} future NA62 physics programme\rev{,} including an increase in beam intensity.
For a better trigger selectivity two systems are under development to extract high level features from the RICH online data stream at full granularity.
A GPU-based system performing a geometrical ring reconstruction algorithm is under finalization in the experimental hall. 
A preliminary neural network model to be implemented on the \lzerotpp FPGA is under development.
They both satisfy the requirements to run online and are examples of distributed trigger computing paradigms with potential benefits for physics discovery.
\acknowledgments
This work has been carried out within the EuroEXA and TEXTAROSSA projects, under grant agreements EU H2020 FP No 754337 and H2020-JTI-EuroHPC-2019-1 No 95683.


\end{document}